\documentclass[12pt]{article}
\usepackage{floatfig,epsfig} 
\DeclareMathAlphabet{\mathsc}{OT1}{cmr}{m}{sc}
\newlength{\dinwidth}                       
\newlength{\dinmargin}                      
\def\lsim{\mathrel{\rlap{\lower4pt\hbox{\hskip1pt$\sim$}}
    \raise1pt\hbox{$<$}}}                
\def\gsim{\mathrel{\rlap{\lower4pt\hbox{\hskip1pt$\sim$}}
    \raise1pt\hbox{$>$}}}                
\newcommand{\apm}{(\alpha_{\mathsc{I\!P}} - 1)}  
\newcommand{\vr}{{\bf r}}
\newcommand{\vb}{{\bf b}}
\newcommand{\e}{{\rm e}}
\newcommand{\df}{{\rm d}}

\begin{document}
\initfloatingfigs

\begin{flushright}
SPhT-96/087\\
Cavendish-HEP-96/10 \\
hep-ph/9607474
\end{flushright}
\vspace*{0.1cm}


\begin{center}  \begin{Large} \begin{bf}
{\boldmath The QCD dipole picture of small-$x$ physics}\\ 
  \end{bf}  \end{Large}
  \vspace*{5mm}
  \begin{large}
R.~Peschanski$^a$, G.~P.~Salam$^b$
\\ 
  \end{large}
\end{center}
$^a$ CEA-Saclay, Service de Physique Th\' eorique, F-91191
 Gif-sur-Yvette Cedex, FRANCE\\
$^b$ Cavendish Laboratory, Cambridge University, Madingley Road,
     Cambridge CB3 0HE, UK  
\begin{quotation}
\noindent
 {\bf Abstract:} The QCD dipole picture of BFKL dynamics provides an
 attractive theoretical approach to the study of the QCD (resummed)
 perturbative expansion of small-$x$ physics and more generally to
 hard high-energy processes. We discuss applications to the
 phenomenology of proton structure functions in the HERA range and to
 the longstanding problem of unitarity corrections, and outline some
 specific predictions of the dipole picture.
\end{quotation}

\section{Introduction}
\sloppy
\begin{floatingfigure}{0.5\textwidth}
\begin{center}
\epsfig{file=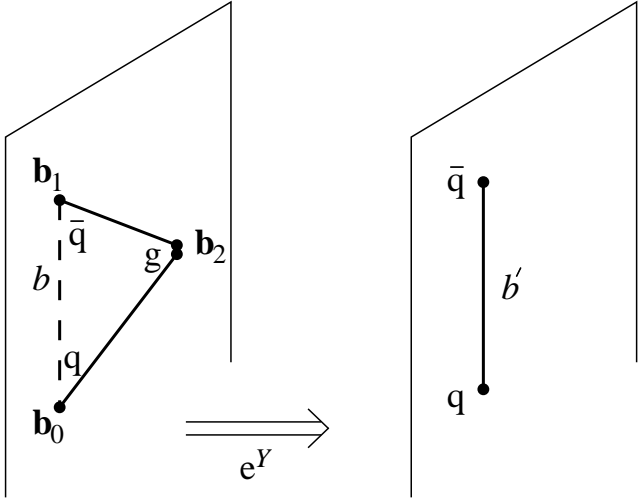, width=0.5\textwidth}
\end{center}
\caption[]{{\it
 Onium-onium interaction at first order via the one-soft-gluon
 component of the onium wavefunction; under the effect of the Lorentz
 boost $\e^Y$, the original ${\bar q} q$ configuration of size $b$ gives
 rise to a soft gluon component, or in the $N_c \rightarrow \infty$
 limit, to two dipoles of sizes $b_{02}$ and $b_{12}$ interacting with
 the other onium of size $b'$. \\ }}
\label{fig:ooint}
\end{floatingfigure}

The dipole formulation \cite{Muel94a,NiZa93a} is an approach to
small-(Bjorken)$x$ physics which for inclusive quantities can be shown
\cite{Muel95} to be equivalent to the BFKL approach \cite{BaLi78}. One
starts with a $q\bar{q}$ state (onium), taken to be heavy enough to
ensure the validity of perturbation theory. The main ingredients of
the dipole picture of BFKL dynamics are the following

 i) Choosing the quantisation in the infinite-momentum frame of the
 onium allows one to select the leading $\alpha \log 1/x$ terms of the
 QCD perturbative expansion of the onium wave-function.
 
 ii) Changing the momentum representation into a mixed one $(\vb,x)$,
 where $\vb$ is the transverse coordinate, amounts to killing the
 contributions of the interference Feynman diagrams in the leading-log
 expansion. This results in a quasi-classical picture of the system of
 quarks and gluons in terms of probability distributions at the
 interaction time.
 
 iii) Finally the $1/N_c$ limit leads to the emergence of a
 representation in terms of independent colourless dipoles, replacing
 the description in terms of soft, coloured gluons. 

To illustrate these properties on a simple example, one constructs the
component of the squared wave function that contains one soft gluon,
as a function of the transverse positions $\vb_0,\vb_1 $ (or impact
parameter) of the onium quark and antiquark and $\vb_2$ of the gluon,
(see fig.~1). In the large-$N_c$ limit, the original colour dipole of
the onium state (of size $b$) effectively becomes two colour dipoles:
one formed by $qg$ (of size $b_{02}$) and the other by $g\bar{q}$ (of
size $b_{12}$). So the addition of a gluon is equivalent to the
branching of one dipole into two, and each of the produced dipoles can
then branch independently --- this leads to a cascade of dipoles
developing when $x$ becomes smaller and smaller, explaining the rise
in the number of dipoles (or gluons) at small $x$.

 To determine the gluon distribution, one must use some probe. One way
 is to measure the interaction cross section with a second onium. The
 evolution equation for the interaction cross-section (see fig. 1) of
 two $q\bar{q}$ states of sizes $b$ and $b'$ is

\begin{eqnarray}
 \frac{\df \sigma(b',b,Y)}{\df Y}
 = \frac{\alpha N_c}{2\pi^2}
   \int \frac{b^2 \df^2 \vb_2}{b_{02}^2 b_{12}^2}
 \big[\sigma(b', b_{02}, Y) \nonumber \\ 
 + \sigma(b', b_{12}, Y) 
	- \sigma(b', b_{01}, Y) \big],
\end{eqnarray}

 \noindent where $Y \simeq \ln 1/x$ is known as the rapidity. 
The solution is

\begin{equation}
 \sigma(b,b',Y) = \frac{8\pi\alpha ^2 bb'}{\sqrt{\pi k Y}}
		\e^{\apm Y - \ln^2(b'/b) / kY}
\label{eq:sigmabb}
\end{equation}

 \noindent with $\apm=(4\ln2)\alpha N_c/\pi $ and $k=\frac {\alpha N_c}\pi
 14\zeta (3)$. Eq.~(\ref{eq:sigmabb}) has some interesting features
 which deserve comment. First it reproduces exactly the high-energy
 ($\simeq$ small-$x$) behaviour associated with the BFKL "hard"
 Pomeron. Second, and more intriguing, a dependence appears on the
 scale-ratio $b'/b$ between the two colliding onia. This is related to
 the property of BFKL dynamics that it ``explores'' a large region in
 the transverse-momentum plane, which is analogous to a classical
 diffusion mechanism.

\section{Structure functions}

 The scale-ratio dependence obtained in formula (2) is of importance
 when considering another type of probe, a photon of virtuality $Q^2$,
 which corresponds on average\cite{NiZa93a} to a transverse size
 $1/Q$. In ref.~\cite{NPR95}, the (theoretical) process of
 deep-inelastic scattering on an onium state has been proposed to
 determine the origin of scaling violations of the structure function
 in the context of BFKL dynamics. Indeed from the viewpoint of the
 dipole picture, scaling violations are induced by a term analogous
 to the scale-ratio in eq.~(\ref{eq:sigmabb}). One gets:

\begin{equation}
 F_2^{onium}  \propto \int^{ }_{ } \frac{d \gamma}{2 i \pi} (bQ)^{2\gamma} 
 \e^{\frac{\alpha N_c}{\pi} \chi(\gamma) \ln\frac{1}{x}} \propto bQ \ 
x^{- \left(\frac{4 \alpha N_c \ln 2}{\pi}\right) }\ 
 \frac {  \exp\left( - \frac{1}{k \ln\frac{1}{x}} \ln^2(bQ)\right)}
{\left(k \ln\frac{1}{x} \right)^{1/2}},
\end{equation}

\begin{figure}[tb]
\begin{center}
\epsfig{file=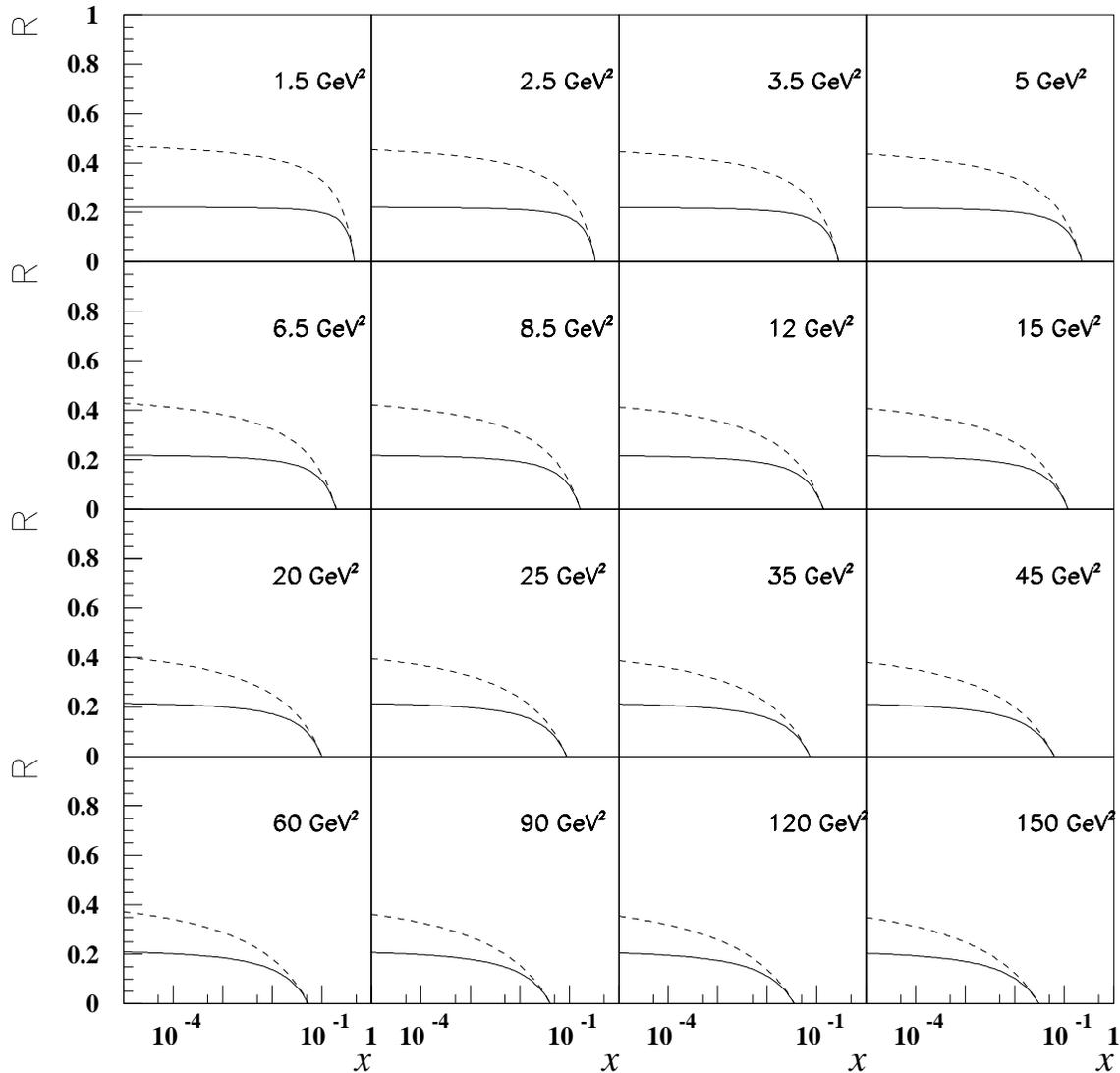, height=0.6\textheight}
\end{center}
\caption[]{\itshape Predictions for the ratio $R \equiv {F_L}/{F_T}$ in
the dipole picture
\cite{NPR95}. The full line describes the prediction based on a fit
to $F_2$ data and, with the same parameters, a determination of the
gluon structure function (not shown). The effect of the $\ln {1 \over
x}$ resummation is seen by comparison with the one-loop approximation
(dotted line). The prediction is significantly lower than the known
DGLAP estimates, e.g.\ \cite{H1}.}
\end{figure}

\noindent where one uses the known BFKL analytic expression for the
Mellin transform of the onium structure function, and $\chi(\gamma)$
is the corresponding kernel\cite{BaLi78}.  This expression leads to an
interesting phenomenological extension to the proton structure
functions, which has the property that it describes the scaling
violations at small-$x$ observed at HERA.

Indeed, assuming $k_T$-factorisation properties\cite{catani} for
high-energy scattering off a proton target, it is possible to extend
the dipole model to deal with deep-inelastic scattering on a proton
target\cite{NPR95}. Starting from formula (3), the Mellin integrand
happens to be multiplied by $ w(\gamma,b;Q_0)$ where $w$ can be
interpreted as the Mellin-transformed probability of finding a dipole
of (small) transverse size $b$ in the proton. $Q_0 >> b^{-1}$ is a
typically non-perturbative proton scale.  Noting that $b$ is a small
but arbitrary factorisation scale, the overall result has to be
$b$-independent, provided it stays in the perturbative region. Hence,
assuming renormalisation group properties to be valid\cite {dejujula},
the $b$ dependence of $w$ has to match the $b^{-2\gamma}$ dependence
in formula (3).  One then writes

\begin{equation}
w(\gamma,b;Q_0) 
=  w(\gamma)\ \left(bQ_0\right)^{2\gamma}. 
\end{equation}

\noindent This yields the final result\cite{NPR95}

\begin{eqnarray}
\left(\begin{array}{c}
F_T \\ F_L \\ F_G \end{array} \right) 
= \frac{2{\alpha} N_c}{\pi} \int^{ }_{ }  \frac{d \gamma}{2 i \pi}  
  \left(\frac{Q^2}{{Q_0}^2}\right) ^{\gamma} 
\e^{\frac{{\alpha} N_c}{\pi} \chi(\gamma) \ln(\frac{1}{x})} 
\left(\begin{array}{c}
h_T \\ h_L \\ 1 \end{array} \right)
\frac{v(\gamma)}{\gamma} w(\gamma)
\end{eqnarray}

\noindent where $F_{T(L)}$ is the structure function corresponding to
transverse(longitudinal) photons and $F_G$ the gluon structure
function.  The known (resummed) coefficient functions
$h_{T,L}(\gamma)$ are given in ref.~\cite{catani}, and the gluon-dipole
coupling $v(\gamma)$ is derived in the second of
refs.~\cite{NPR95}. It is interesting that these formulae give a good
fit of the HERA data on $F_2 =F_T + F_L$ in the small-$x$ range and in
a large domain of $Q^2 \le 150\ GeV^2.$ Moreover it leads to a gluon
structure function in agreement with the H1 determination based on the
next-leading order DGLAP evolution\cite{H1}. Note also that the ratios
${F_G}/{F_2}$ and $R
\equiv {F_L}/{F_T}$ are independent of the non-perturbative function
$w(\gamma).$ In relation to this a remark is in order for the future
prospects of experimentation at HERA: As shown in fig.2, the
predictions for $R$ are rather low  ($R<2/9$) which
appears to be in contradiction with the phenomenological estimate\cite
{H1} based on the renormalisation group evolution for $F_L.$ Indeed,
as shown in fig.~2, the resummation of the leading $\alpha \log 1/x$
terms of the QCD perturbative expansion is crucial for obtaining the
final prediction. This may give a hint for an experimental
discrimination of DGLAP versus BFKL evolution equations which is
difficult to achieve from the study of $F_2$ and $F_G$ alone.

Another series of interesting phenomenological results apply to hard
diffraction at HERA. The QCD dipole picture leads to two distinct
dynamical components of diffraction by a virtual photon. One
component, dominant at large diffractively produced masses, is
analogous to the triple-(hard)Pomeron coupling and can be explicitly
derived from the inelastic interaction of dipoles from both the photon
and the proton sides \cite{BP95}; A second component results from the
{\it quasi-elastic} interaction of the primary dipole coupled to the
photon to the proton target and is dominant at smaller diffractive
masses \cite{BP96}; The quantitative predictions from these two
components are strongly correlated with the fits for $F_2$, giving a
nice interrelation between the different aspects of deep-inelastic
processes at HERA and the possibility to rely on perturbative QCD to
get a coherent description for them.

\section{Unitarity corrections}


 When the centre-of-mass energy becomes very high, the BFKL equation
 yields a scattering amplitude which violates the unitarity bound, or
 equivalently conservation of probability. The dipole formulation
 offers a well-defined way of alleviating the problem. One considers
 the scattering of two onia in the centre of mass frame.
 Schematically the scattering amplitude
 is just related to the probability that there will be an interaction
 between a parton in one onium and a parton in the other. The usual
 small-$x$ growth of the cross section relies on the idea that the
 interaction cross section is proportional essentially to the product
 of the number of partons in each onium. This is only valid when the
 overall likelihood of an interaction is low. When there are many
 partons in each onium, multiple interactions become
 common\cite{GrLR83}, and the interaction probability then depends on
 the details of how the partons are distributed in transverse position
 (for example if they are clumped together, then multiple interactions
 are much more likely than if they are uniformally spread out).  These
 multiple-scattering corrections are equivalent to multiple
 $t$-channel pomeron exchange diagrams\cite{Muel94a,Sala95}.

\sloppy

\begin{floatingfigure}{0.55\textwidth}
\begin{center}
\epsfig{file=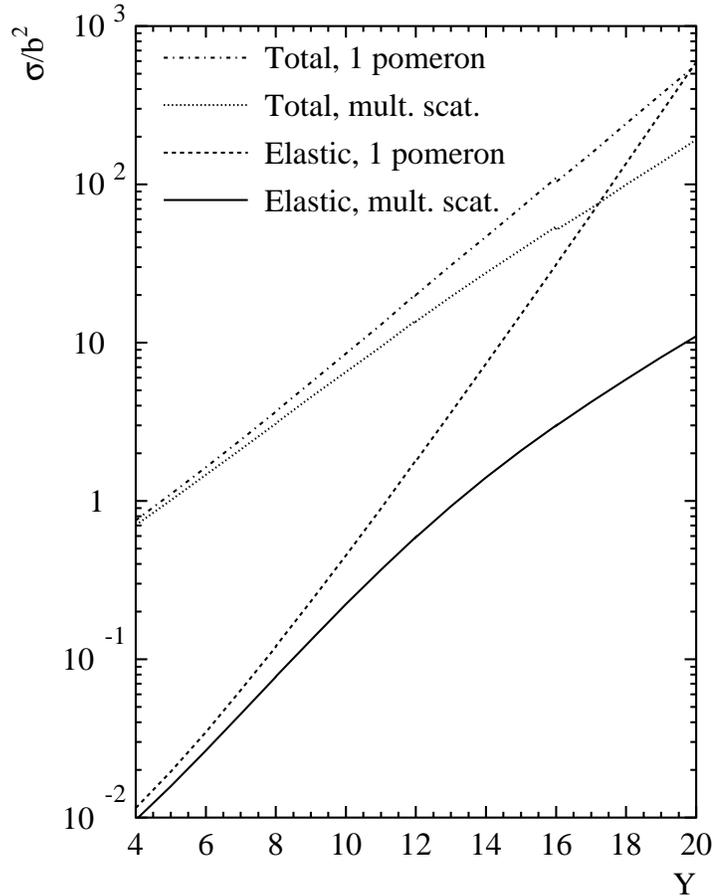, width=0.55\textwidth}
\end{center}
\caption[]{{\it
 The elastic and total cross
 sections for onium-onium scattering, as a function of rapidity,
 showing both the one-pomeron approximation and the results including
 multiple-scattering corrections.
  }}
\label{fig:sigeltot}
\end{floatingfigure}

 To obtain the probabilities of different gluon distributions inside
 the onium, one can use OEDIPUS 
 (Onium Evolution, Dipole Interaction and  
 Perturbative Unitarisation Software) \cite{Sala96a}. This
 simulates the small-$x$ dipole branching producing random dipole
 configurations with the correct weights. It determines the
 interaction (both with and without multiple-scattering corrections)
 between pairs of these random configurations and then averages over
 the configurations. It is important that one averages over
 configurations only {\em after} taking into account multiple
 interactions --- doing the averaging before taking into account the
 multiple interactions (the eikonal approximation) tends to wash out
 the correlations between gluons, and causes one to underestimate the
 point where corrections set in by up to two orders of magnitude in
 $x$.


 The results \cite{Sala95} are shown in figure~\ref{fig:sigeltot}. The
 rapidity $Y$ corresponds roughly to $\ln 1/x$, and $b$ is the onium
 size. The most striking point is that corrections to the total cross
 section set in very slowly, whereas the elastic cross section is
 subject to very strong modifications. The reason is that the total
 cross section is proportional to the integral over impact parameter,
 $\vr$, of the amplitude $F(\vr)$, whereas the elastic cross section
 is proportional to the integral of the square of the amplitude:

\begin{equation}
 \sigma_{{\rm tot}}(Y) = 2\int \df^2\vr \: F(\vr), \qquad
 \sigma_{{\rm el}}(Y) = \int \df^2\vr \: |F (\vr)|^2.
\end{equation}


 \noindent Because of BFKL diffusion, for moderate $r=|\vr|$ the leading
 dependence of the amplitude is $F(r) \sim 1/r^2$. Therefore the
 elastic cross section is dominated by small impact-parameters, where
 the amplitude is large and there are strong multiple-scattering
 corrections. The total cross section comes from a wide range of $r$,
 where the amplitude will on average be smaller, and so the
 corrections are less important. Effectively the total cross section
 carries on growing through an increase in area of interaction. More
 details can be found in \cite{Sala95}

 Onium-onium scattering is a good theoretical laboratory because it
 ensures that it is safe to use perturbative QCD. For DIS one can
 expect two major qualitative differences: (a) infra-red effects will
 constrain the maximum size of the dipoles, limiting the growth of the
 total cross section, and altering the balance between total and
 elastic cross sections; (b) the presence of two different scales
 means that different kinds of dipole configurations will dominate the
 scattering, tending to reduce the multiple-interaction effects.


\section*{Acknowledgements}
 One of us (GPS) is grateful to B.~R.~Webber and A.~H.~Mueller for
 many helpful discussion and to the UK PPARC and the EU Programme
 ``Human Capital and Mobility'', Network ``Physics at High Energy
 Colliders'', contract CHRX-CT93-0357 (DG 12 COMA), for financial
 support. R.P. thanks A.~Bialas, H.~Navelet, Ch.~Royon and S.~Wallon
 for a fruitful collaboration.

\end{document}